\documentclass[conference]{IEEEtran}
\usepackage{textcomp}
\usepackage{hyperref}
\usepackage{amsmath}
\usepackage{float}
\usepackage[justification=raggedright,singlelinecheck=false]{caption}

\makeatletter

\def\ps@IEEEtitlepagestyle{%
  \def\@oddfoot{\mycopyrightnotice}%
  \def\@evenfoot{}%
}
\def\mycopyrightnotice{%
  {\footnotesize 978-1-6654-3835-5/25/\$31.00~\copyright~2025 IEEE\hfill}% <--- Change here
  \gdef\mycopyrightnotice{}
}

\usepackage{blindtext}
\usepackage{eso-pic}
\IEEEoverridecommandlockouts
\usepackage{cite}
\usepackage{amsmath,amssymb,amsfonts}
\usepackage{algorithmic}
\usepackage{graphicx}
\usepackage{textcomp}
\usepackage{xcolor}
\def\BibTeX{{\rm B\kern-.05em{\sc i\kern-.025em b}\kern-.08em
    T\kern-.1667em\lower.7ex\hbox{E}\kern-.125emX}}
    
\usepackage{eso-pic}
\newcommand\AtPageUpperMyright[1]{\AtPageUpperLeft{%
 \put(\LenToUnit{0.17\paperwidth},\LenToUnit{-2cm}){%
     \parbox{0.9\textwidth}{\raggedleft\fontsize{8}{11}\selectfont #1}}%
 }}%
\newcommand{\conf}[1]{%
\AddToShipoutPictureBG*{%
\AtPageUpperMyright{#1}
}
}

\begin{document}
\title{\vspace*{1cm} Ionospheric VLF Radio Reflection Analysis System
}

\author{\IEEEauthorblockN{George Dan Chita}\\
\textit{"Emil Racovita" National College}\\
Cluj-Napoca, Romania\\
george.chita48@gmail.com}

\maketitle
\conf{\textit{  V. International Conference on Electrical, Computer and Energy Technologies (ICECET 2025) \\ 
3-6 July 2025, Paris-France}}
\vspace{-30pt}
\begin{abstract}
Historically, solar flare detection has been dependent on methods that require the presence of expensive satellites or other Earth-based costly equipment. In this paper, we propose a cost-effective, terrestrial alternative that enables reliable solar flare detection. We will discuss the design, practical implementation, and demonstration of a monitoring system for VLF (Very Low Frequency, 3 kHz–30 kHz) radio signals transmitted by stations located thousands of kilometers away. This frequency range was selected because VLF radio waves are efficiently reflected by the lower ionospheric layers, and any changes in these layers lead to corresponding variations in the received signal. We have used a magnetic loop antenna for signal reception along with a low noise amplifier, analog-to-digital converter, and spectrum analyzer. Data was collected over the course of a few days, from a remote location with minimal electromagnetic interference. From the viewpoint of cost, simplicity and accessibility, this method far surpasses traditional methods of detecting solar flares, using satellites, telescopes, etc.\\
\end{abstract}

%\copyrightnotice{979-8-3315-3559-9/25/$31.00 ©2025 IEEE}

\section{Introduction}

Solar flare detection has traditionally relied on satellite-based methods and equipment, making it costly and inaccessible to most. This paper proposes an alternative detection approach that is Earth-based, more affordable, and thus widely accessible.

Intense solar flares emit strong X-rays, which are detected by artificial satellites approximately 8 minutes after the flare begins. These flares also cause a surge in solar wind intensity and increase the kinetic energy of its charged particles. Upon reaching Earth's vicinity, after dozens of hours, the solar wind interacts violently with the planet’s magnetic field, triggering intense geomagnetic storms. Such storms can severely impact telecommunications infrastructure and power grids, causing major damage. (e.g. The Great Quebec Blackout \cite{quebecBlackout2021}).

This study explores an alternative method for detecting solar flares by analyzing VLF (Very Low Frequency, 3 kHz–30 kHz) radio signals transmitted by distant stations. ELF/VLF waves are especially valuable due to their reflectivity in the Earth's ionospheric D region (60–90 km altitude) (See Fig. 1 for ionosphere layers). 

\begin{figure}
    \centering
    \includegraphics[width=0.85\linewidth]{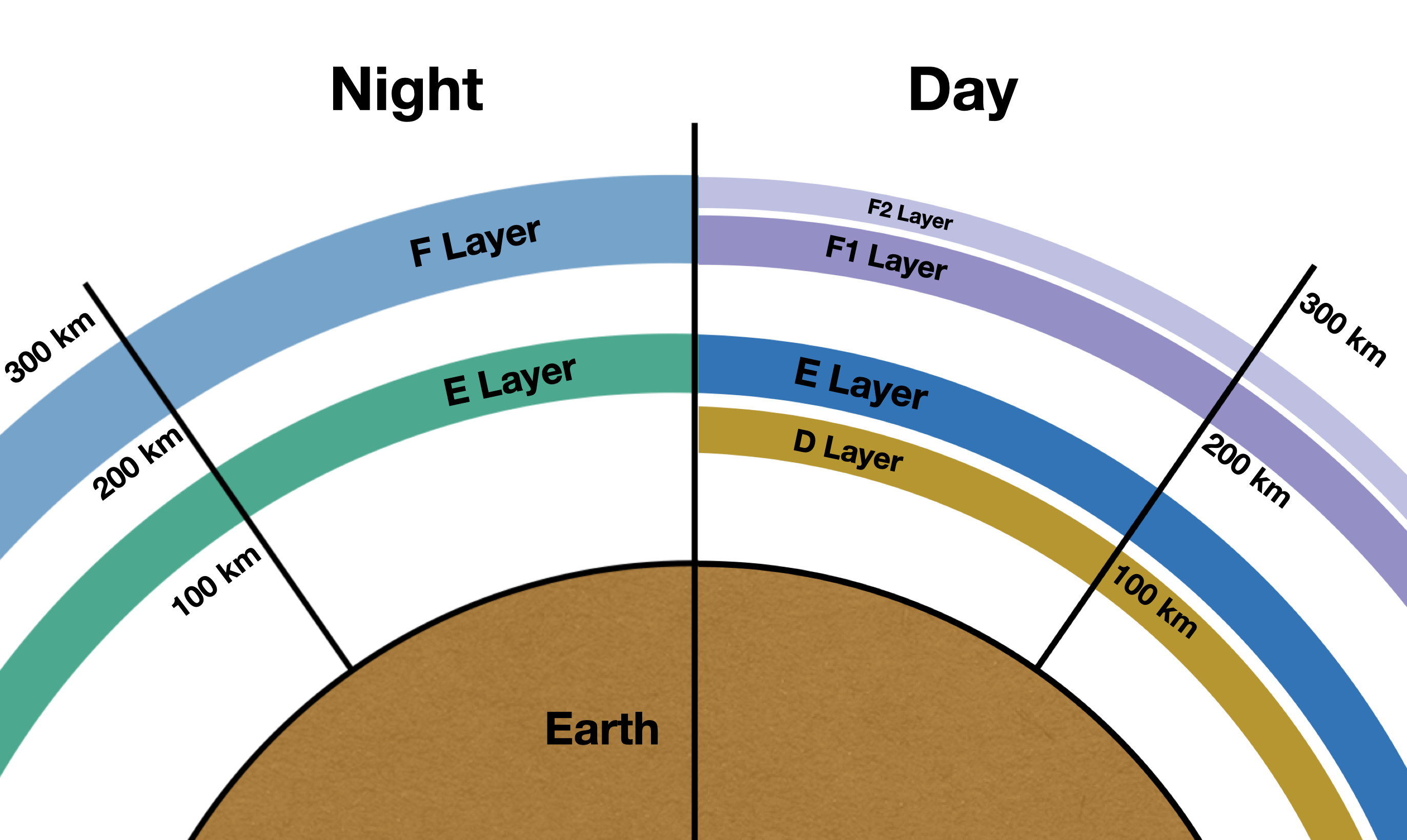}
    \caption{\fontsize{8}{10}\selectfont Ionosphere's Layers Day/Night}
    \label{fig:enter-label}
\end{figure}

They propagate efficiently within the Earth-ionosphere waveguide, covering vast distances. This characteristic allows ELF/VLF waves to serve as a remote sensing tool for monitoring the D region, which is influenced by solar activity, atmospheric discharges, electron precipitation from radiation belts, cosmic gamma rays, and seismic events\cite{vlfstanford}.

At the reception point, the measured signal is the result of interference between the ground wave and sky wave \cite{UnderstandingHFPropagation} (See Fig. 2). These two can combine constructively or destructively and changes in the ionosphere thickness can result in either a rise or a drop in signal, respectively.

\begin{figure}[H]
    \centering
    \includegraphics[width=0.75\linewidth]{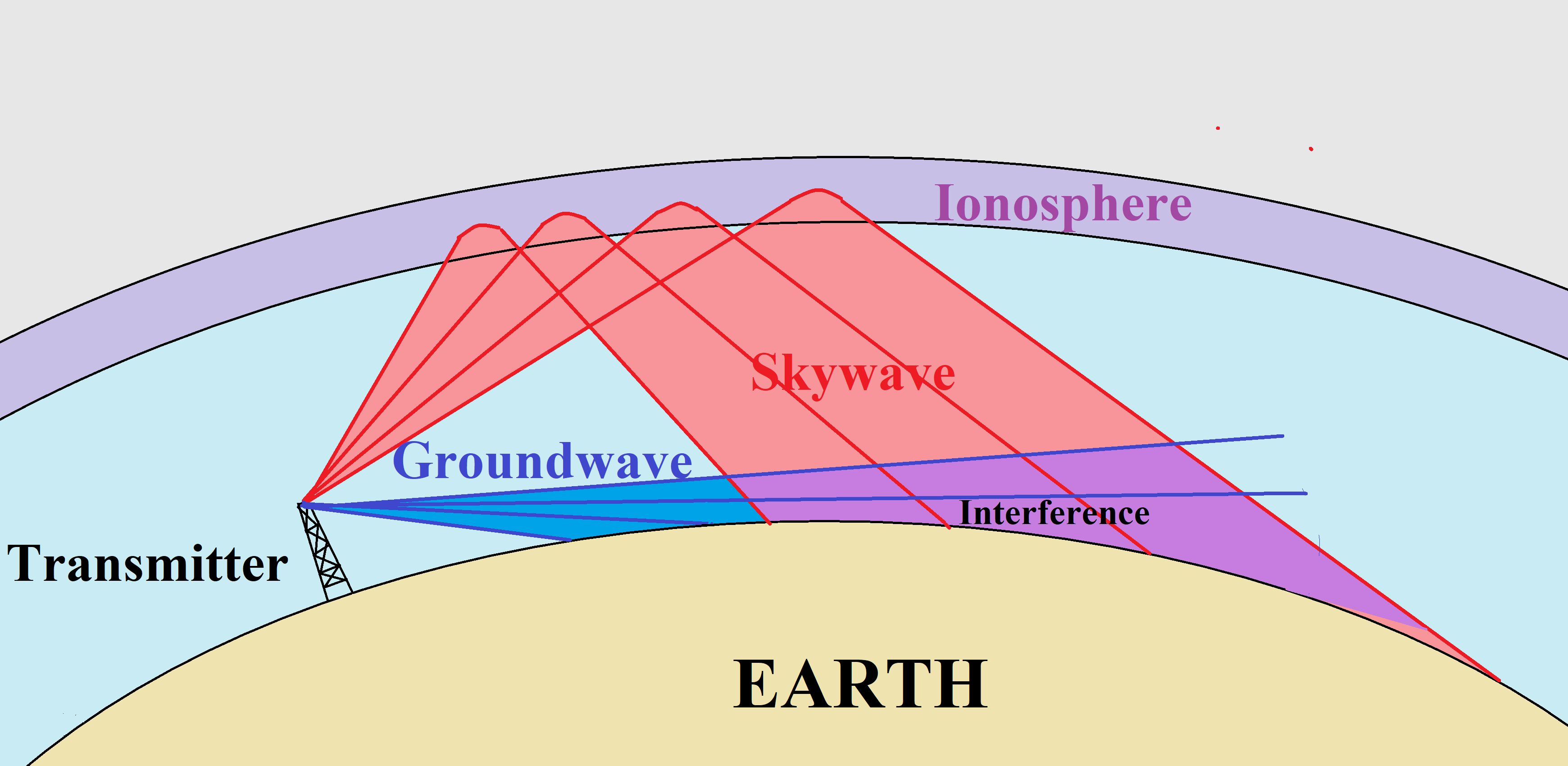}
    \caption{\fontsize{8}{10}\selectfont Ground and Sky Wave Propagation}
    \label{fig:enter-label}
\end{figure}

\textbf{Advantages over Traditional Methods} : The most precise, quantitative way of detecting solar flares and measuring X-ray flux intensity is by using X-ray photometers on satellites (ex: GOES \cite{laspGoesXrayFlux}). This method allows for accurate detection of particles coming from the Sun, such as X-ray photons, and displaying this reading in W/m². However, satellite systems are costly and overly complicated, for most research institutions. Terrestrial alternatives include detection of solar radio bursts via radio telescopes — effective but similarly expensive — H-alpha solar imaging, neutron monitors, and indirect methods such as magnetic sudden impulse (SI) detection using ground magnetometers. Of these, H-alpha imaging is limited by weather and daylight conditions. In contrast, our proposed method offers a low-cost, easily deployable solution capable of operating continuously, including under overcast skies or without direct solar visibility. Nonetheless, sources of interference such as lightning activity and nearby electronic devices may impact data quality, necessitating deployment in electromagnetically quiet, remote locations.

\section{Methods}
\subsection{Equipment}

\subsubsection{Receiving Loop Antenna}
We chose to use a magnetic antenna for the purposes of this project, as an electrical one would require a wire length that is at least a multiple of $\lambda/4$, which in this case would be around 3 km, an unrealistic option.\\
\\
\textbf{Antenna Housing and Shielding} : The antenna housing was built using a 35 mm diameter, 1 mm thick copper pipe, which provided both structural support for the windings and electrostatic shielding against noise from stray electric fields. The pipe was assembled and soldered into a square frame with a side length of 850 mm, resulting in an approximate surface area of 0.72 m². To simplify the winding process, one side of the pipe was cut along the loop’s plane, granting internal access for feeding the coil wires. Importantly, one corner of the square frame was not electrically connected to avoid creating a conductive loop that could short-circuit the induced signal. Instead, a 3D-printed insulating coupler was used, also serving as a support for signal wire connectors. See Figs. 3-4 :\\
\begin{figure}[!h]
    \centering
    \includegraphics[width=0.6\linewidth]{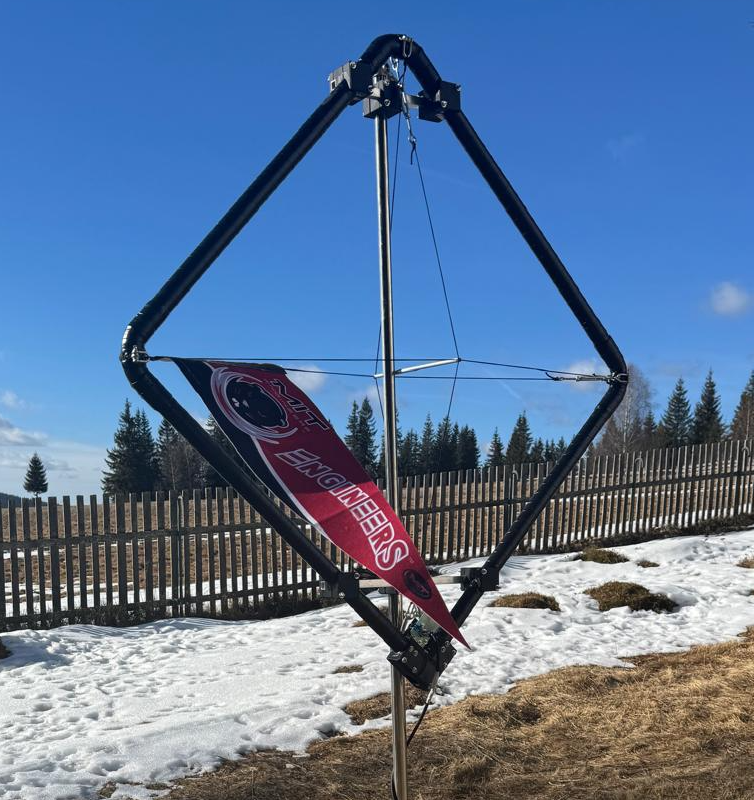}
    \caption{\fontsize{8}{10}\selectfont Magnetic Loop Antenna}
    \label{fig:enter-label}
\end{figure}
\\

\begin{figure}
    \centering
    \includegraphics[width=0.7\linewidth]{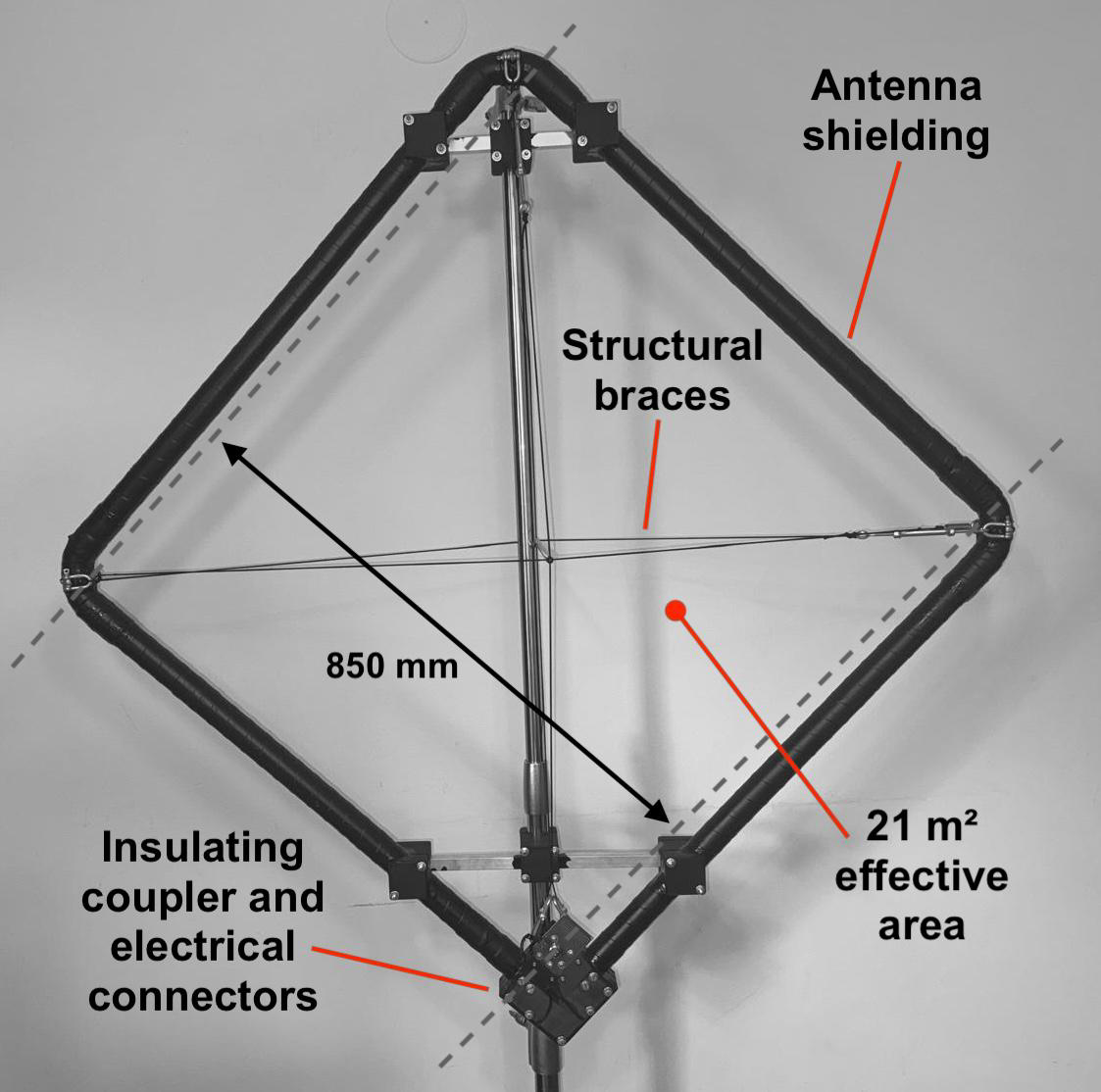}
    \caption{\fontsize{8}{10}\selectfont Magnetic Loop Antenna}
    \label{fig:enter-label}
\end{figure}

\textbf{Coil Windings} : A 29-turn coil was wound inside the housing using 1.5 mm² solid-core copper wire (H07V-U). To ensure sufficient spacing between adjacent windings, thick insulation was chosen, which also helped reduce total capacitance and, as a result, significantly increased the antenna's resonant frequency. We aimed to avoid a resonance frequency that was close to the frequencies we intended to measure, in order to minimize irregular signal response across the radio band.\\

The resonance frequency of the antenna was determined by sweeping through all frequencies with a signal generator and observing for a peak in amplitude and no phase shift between voltage and current through the circuit (See Fig. 7). It was found to be 113.0 kHz, well above our target reception frequencies.\\

\vspace{+10pt}

\textbf{LCR Measurements and Simplified Model}\\

For impedance matching with the low noise amplifier, we created a simplified model (See Fig. 5) and measured parameters such as R, L, and C in order to accurately determine the antenna's impedance. 

\begin{figure}[H]
    \centering
    \includegraphics[width=0.5\linewidth]{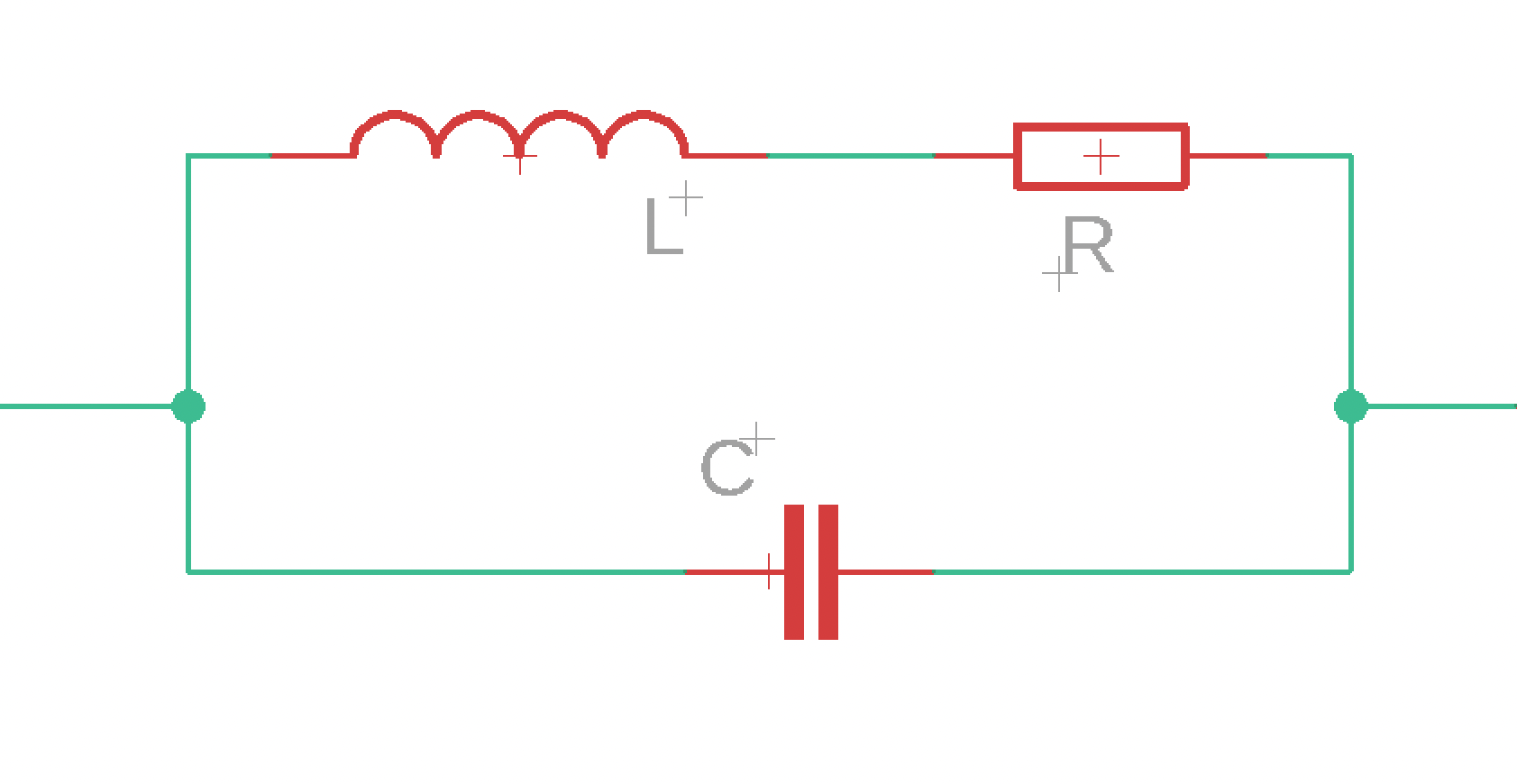}
    \caption{\fontsize{8}{10}\selectfont Simplified Model of Magnetic Loop Antenna}
    \label{fig:enter-label}
\end{figure}

The antenna can be simply modeled as an inductor in series with a resistor, all in parallel with a capacitor (See Fig 5). L represents the inductance of the coil, R the resistance of the wire and C, the parasitic capacitance between windings. The resonance frequency appears when the imaginary part of the impedance:

% Requires: \usepackage{amsmath}
\begin{equation}
    \scriptsize
    \tilde{Z} = \left( \frac{1}{R^2 + \left( wL - \frac{1}{wC} \right)^2} \right) 
        \left( \frac{R}{w^2 C^2} + j\left( \frac{L}{C} \left( wL - \frac{1}{wC} \right) + \frac{R^2}{wC} \right) \right)
    \label{eq:complex_impedance}
\end{equation}

drops to 0 (then voltage and current are in phase), at:

\begin{equation}
    w_{res}=\sqrt{\frac{L - R^2 C}{L^2 C}}
    \label{eq:placeholder_label}
\end{equation}

We cannot use an LCR meter to measure these values as we cannot isolate the individual components of the circuit. A resistor was added in series with the antenna, and voltage and current were measured in the circuit by measuring the voltage on R1 (the added series resistor) and C (voltage across C is across the entire antenna). Using an Ohmmeter we measured  R1=10.43$\Omega$ (R1 is not part of the antenna), R=1.18$\Omega$ and using a signal generator we found $\mathcal{f}_{rez}=$113kHz.\\

Another measurement was done at $\mathcal{f}=$10kHz (See Fig. 6) and $U_{R1}=$264mV, $U_{C}=$3.61V ($U_{C}$ is across the antenna). $\Rightarrow$ \begin{equation}
    Z_{10\text{kHz}} = \frac{U_{C} \cdot R1}{U_{R1}} = 142.23 \approx 142 \, \Omega
    \label{eq:1}
\end{equation}

\begin{figure}[H]
    \centering
    \begin{minipage}[t]{0.48\linewidth}
        \centering
        \includegraphics[width=\linewidth]{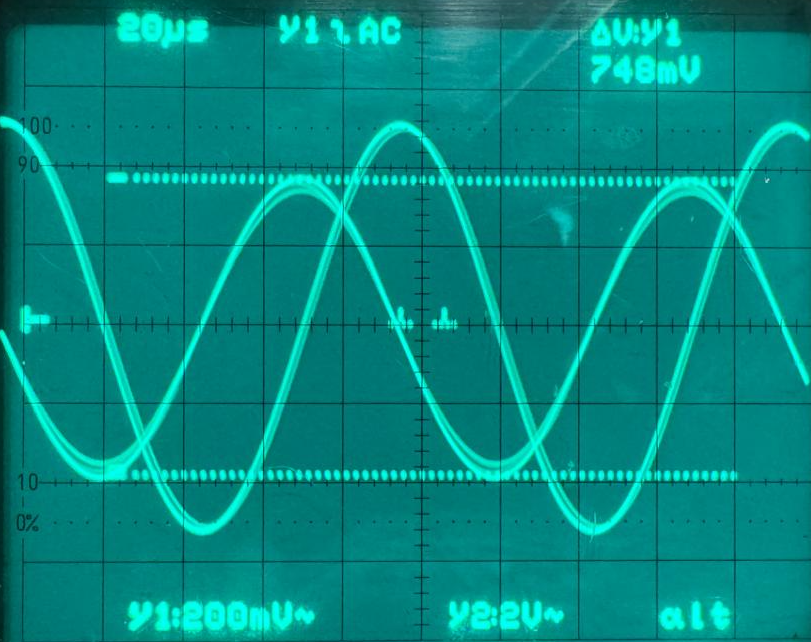}
        \caption{\fontsize{8}{10}\selectfont Measurement at 10kHz}
        \label{fig:image5}
    \end{minipage}
    \hfill
    \begin{minipage}[t]{0.48\linewidth}
        \centering
        \includegraphics[width=\linewidth]{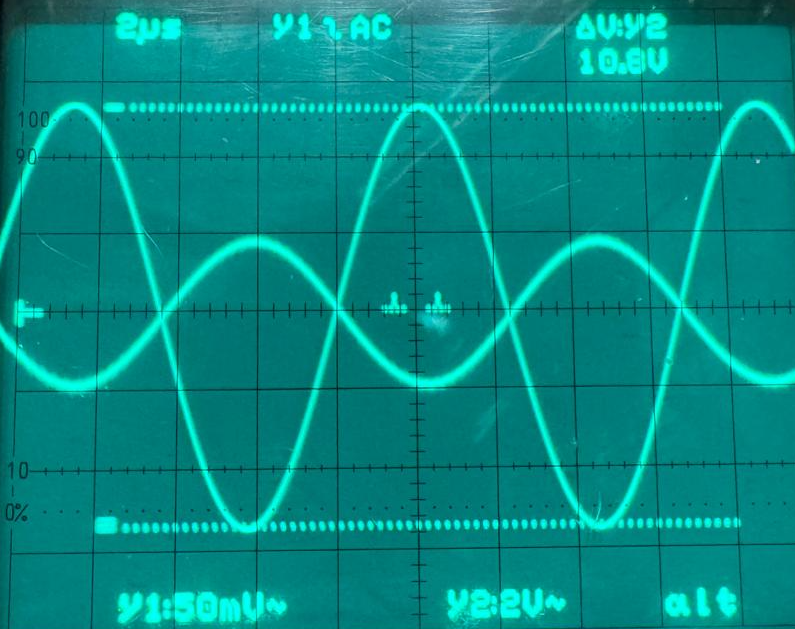}
        \caption{Voltage and Current through Antenna at Resonance}
        \label{fig:image2}
    \end{minipage}
\end{figure}

We observe that the value of R is very small, especially compared to the impedance of the coil at high frequencies. To simplify calculations, we will ignore it when determining the impedance of the antenna, treating it as a parallel LC circuit.
Having these measurements, we can use equations (1) and (2) to get an equation for the inductance of the antenna: 

\begin{equation}
    L = Z \left( \frac{1}{w} - \frac{w}{(w_{\text{res}})^2} \left(1 - \frac{R}{Z_{rez}} \right) \right)
    \label{eq:example_label}
\end{equation}
where:
\begin{equation}
    Z_{\text{res}} = \frac{U_{\text{Cres}} \cdot R1}{U_{\text{R1res}}} = 1217.77 \approx 1220\Omega
    \label{eq:Z_rez}
\end{equation}

\begin{equation}
    L = 2.25 mH\label{eq:example_equation}
\end{equation}
\begin{equation}
    C = 0.88 \, \text{nF} \label{eq:capacitance_value}
\end{equation}

\subsubsection{Mounting and Protection}

The assembly was covered with 38 mm wide self-fusing silicone tape to protect against water and harsh weather conditions. It was then mounted onto a speaker stand using 18 3D-printed brackets and two aluminum square beams.\\

\subsubsection{Low-Noise Signal Amplifier}

The OPAX134 \cite{tiOPA2134} operational amplifier [10] was chosen for its ultra-low distortion (0.00008\%), low noise (8 $\mathrm{nV}/\sqrt{\mathrm{Hz}}$), and effectiveness in audio-frequency amplification. The OPA2134PA package was used, with its two internal amplifiers configured in parallel to further minimize noise and enhance the signal-to-noise ratio.

In the circuit shown below (Figure 8), resistors pairs R5,R4 and R1,R6 define the gain of the two amplifiers using the manufacturer's formula: \\

$$G_1 = 1 + \frac{R_4}{R_5}, \quad G_2 = 1 + \frac{R_1}{R_6}$$

Capacitors C1, C2, C3, and C4 serve as coupling capacitors, while potentiometers R11, R10, and R9 are used to adjust supply voltage levels and input impedance. To reduce circuit noise and minimize parasitic inductance, film resistors and primarily SMD components were used, helping to shorten trace lengths.

\begin{figure}[H]
    \centering
    \includegraphics[width=1\linewidth]{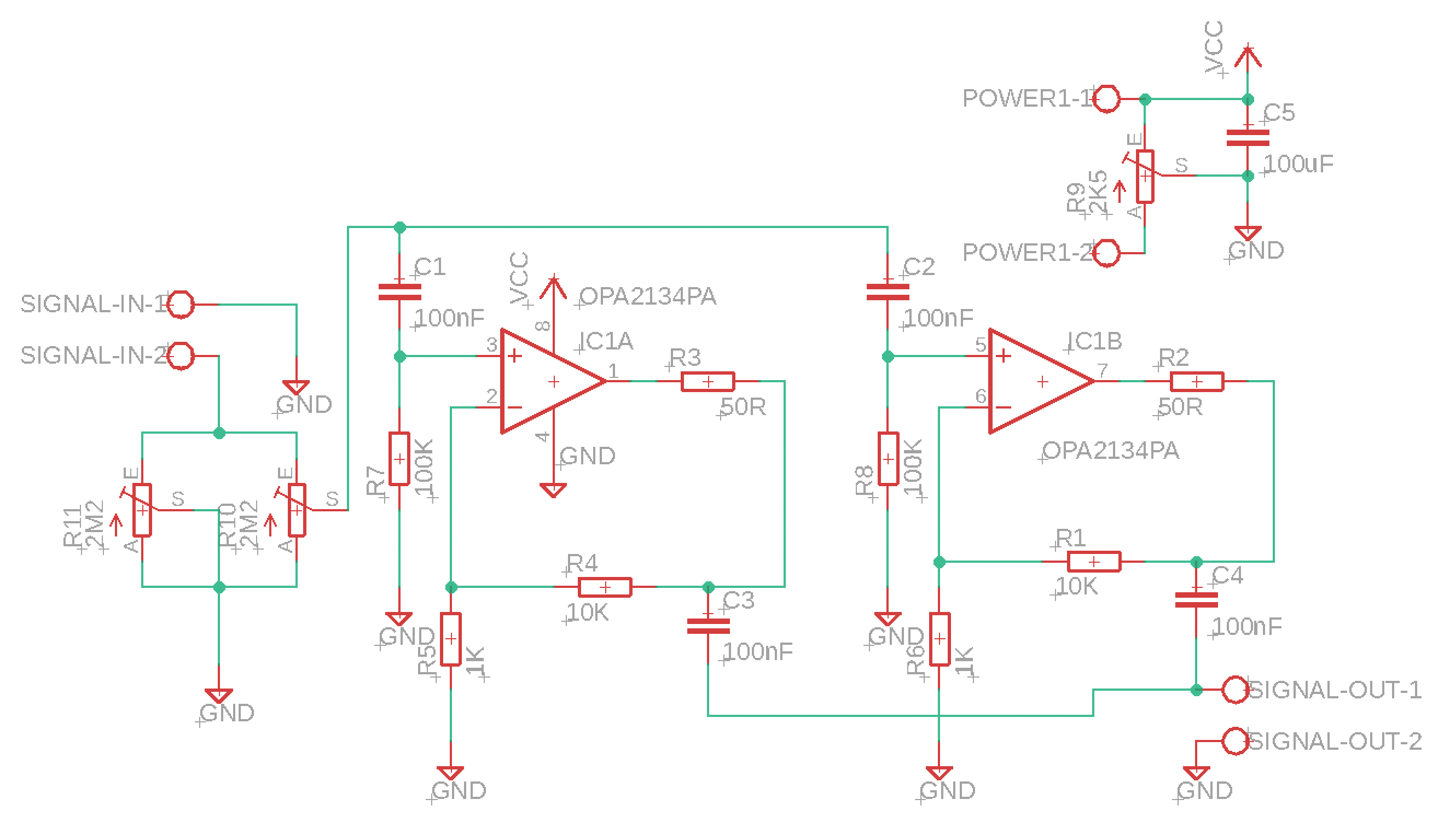}
    \caption{\fontsize{8}{10}\selectfont Low Noise Amplifier Schematic Diagram}
    \label{fig:enter-label}
\end{figure}

\subsubsection{Analog to Digital Converter}
A Steinberg UR22C \cite{UR22C} external sound card was used to introduce a second amplification stage, enhance signal reception, and mainly function as a digital signal processor. Additionally, it provides a wider reception band (0-96kHz) with a sampling rate of 192 kHz (double the reception band width).

The signal output from the amplifier was connected to the sound card input using a shielded microphone-audio jack cable. The sound card was then connected to the laptop, where the signal analysis will be performed.\\

\subsubsection{Laptop Charging}

We found that the Laptop Charger having a switch-mode power supply introduced an unacceptable amount of noise mainly around 50-70kHz but spreading to most other frequencies, drowning many radio signals. We had to build a charging circuit that would not introduce high frequency noise but while still being supplied by a standard wall outlet to ensure continuous operation. A 15V toroidal transformer was used in combination with a simple full bridge rectifier and filter capacitor. The circuit was calculated to output 18-20 V depending on the outlet voltage and load. See schematic in Fig. 9 :

\begin{figure}[H]
    \centering
    \includegraphics[width=1\linewidth]{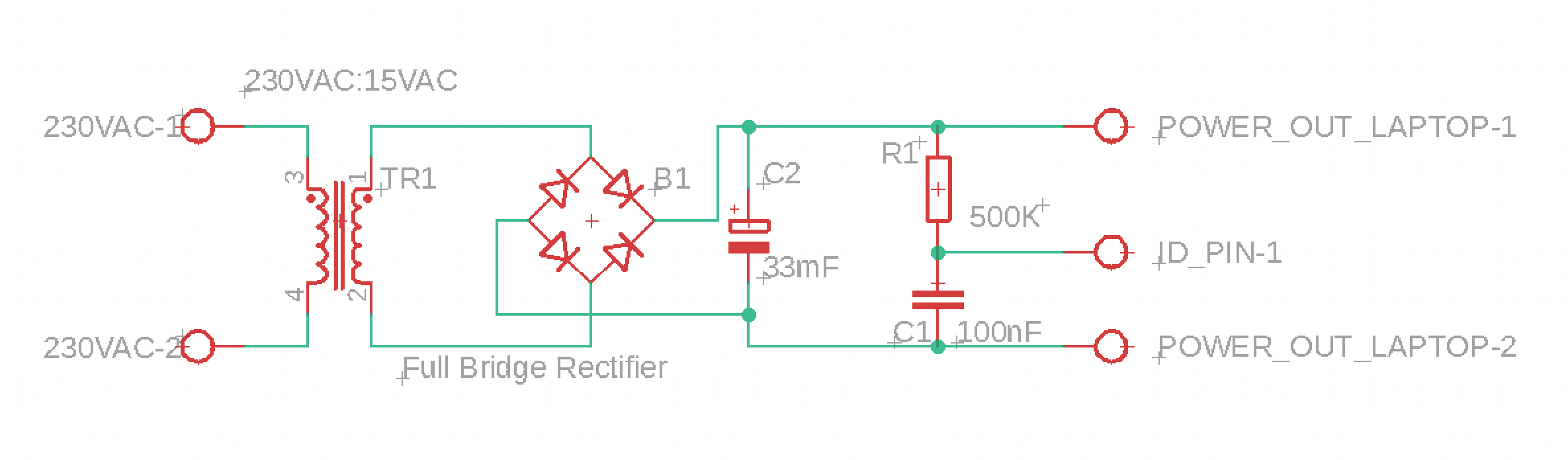}
    \caption{\fontsize{8}{10}\selectfont Laptop Charging Circuit}
    \label{fig:enter-label}
\end{figure}

\subsubsection{Circuit and Cable Shielding}

All circuits were enclosed in conductive casings and placed in a large aluminum box that was connected to ground. All cables used were shielded audio cables, mainly  HELUKABLE
400080 (2×0.5 mm² See datasheet here \cite{Helusound} ). All wire shielding layers were left floating.\\

\subsubsection{Complete System Assembly} See Figs. 10-11 : \\

\begin{figure}[H]
    \centering
    \includegraphics[width=0.75\linewidth]{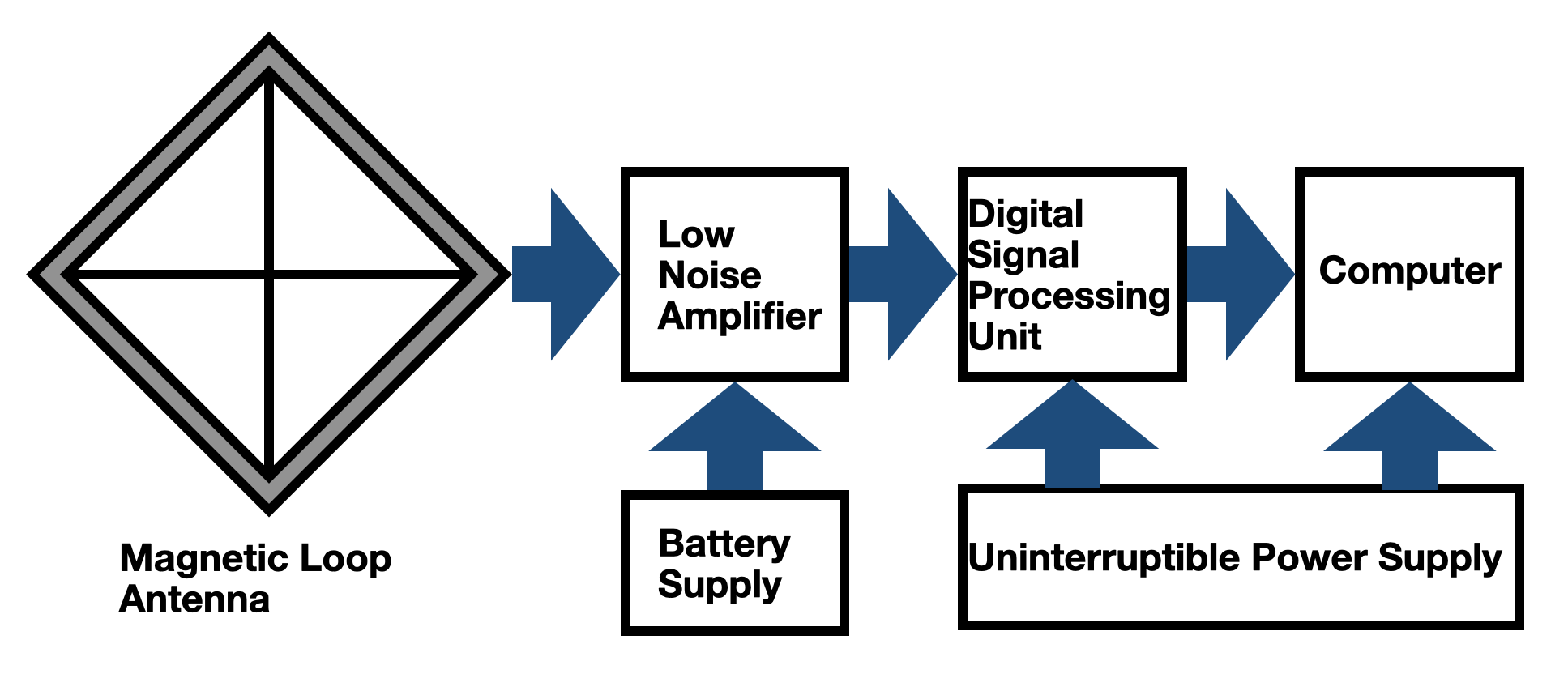}
    \caption{\fontsize{8}{10}\selectfont Block Diagram}
    \label{fig:enter-label}
\end{figure}

\begin{figure}[H]
    \centering
    \includegraphics[width=0.9\linewidth]{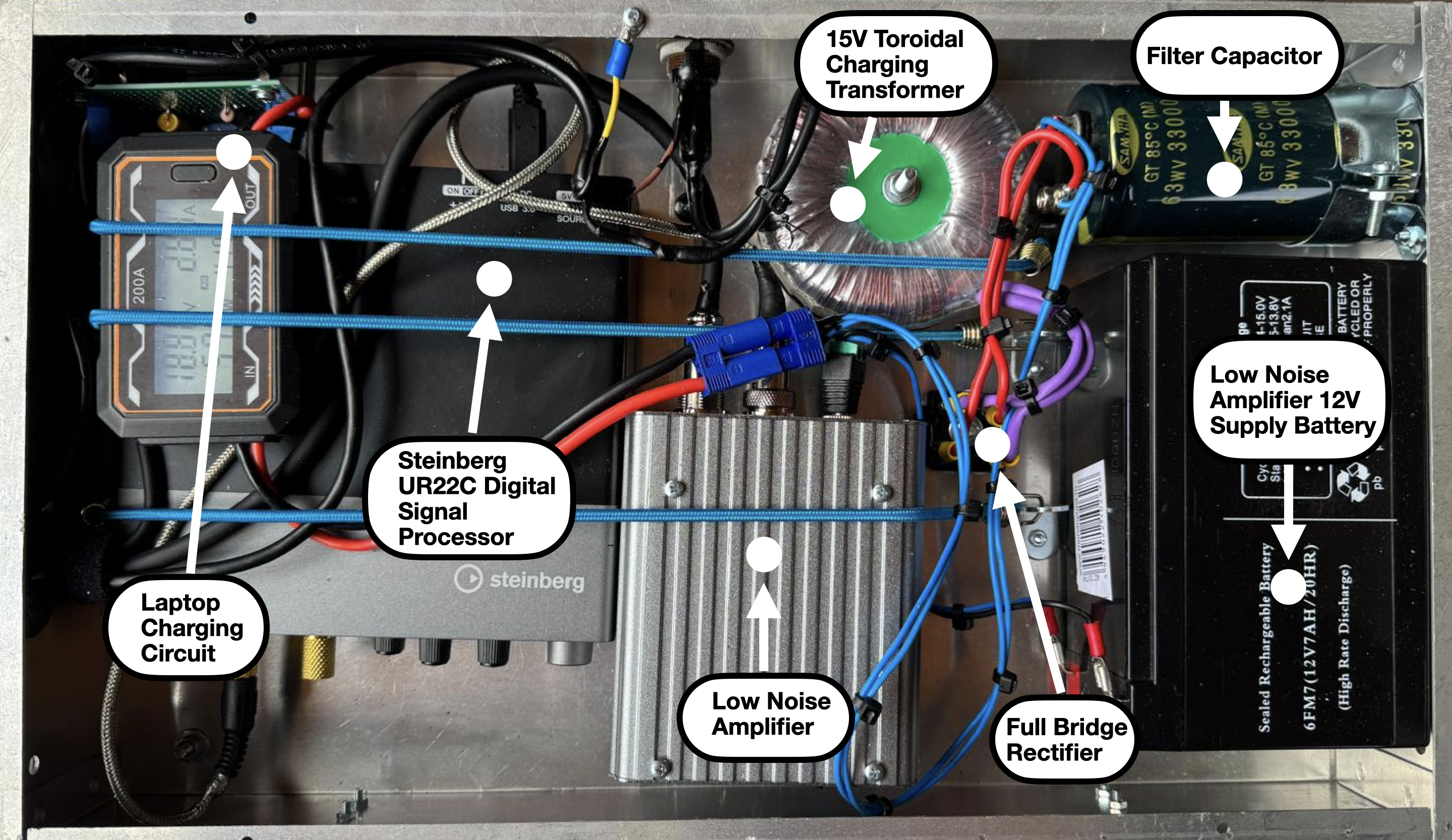}
    \caption{\fontsize{8}{10}\selectfont Complete Assembly}
    \label{fig:enter-label}
\end{figure}

\subsection{Measurements}

\subsubsection{Digital Signal Analysis} We employed Spectrum Lab \cite{SpectrumLab} on a HP Elitebook Laptop to capture, analyze and record data from the antenna. Results were periodically saved in a .txt file which was then imported into Microsoft Excel in order to plot the graphs of amplitude versus time.

\textbf{Settings} : We found the settings depicted in Fig. 12 to bring the best results. The higher FFT setting and averaging, led to a decreased noise background while still allowing sudden change to be detected in the received signal.\\

\begin{figure}[H]
    \centering
    \includegraphics[width=0.8\linewidth]{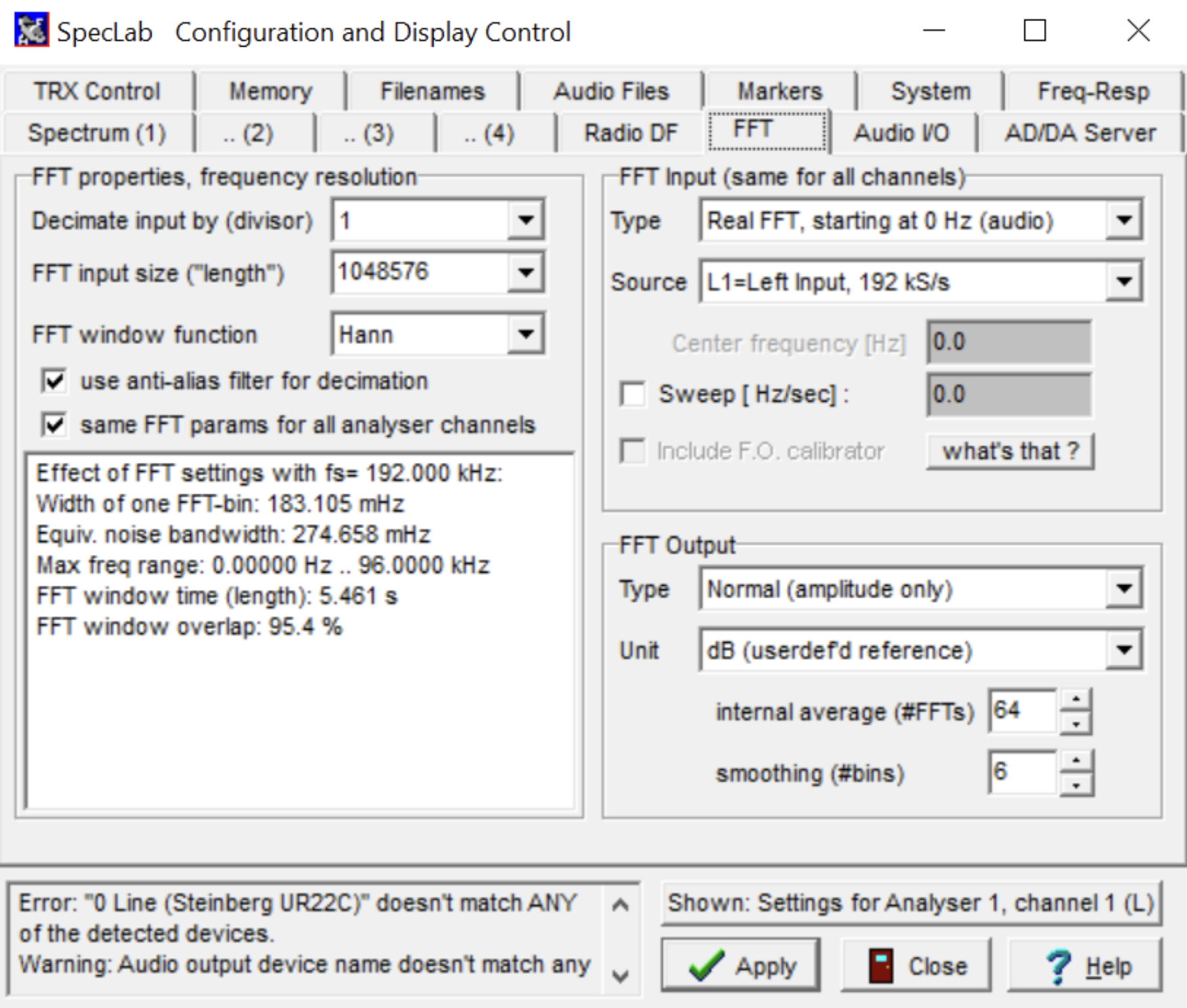}
    \caption{\fontsize{8}{10}\selectfont Optimal Settings}
    \label{fig:enter-label}
\end{figure}

\textbf{Chosen Radio Stations} : Radio transmitters based in Europe were chosen so they would be within a 2000 km radius, resulting in a stronger signal, but also preventing multiple reflections of the ionosphere leading to irregularities and poor signal quality.
Besides this, radio stations were picked to be roughly collinear as our antenna is directional and needs to be oriented along the direction of the transmitter.

See Table I for the list of selected VLF stations and Fig. 13 for the location map of the transmitters.

\begin{figure}[H]
    \centering
    {\fontsize{8}{10}\selectfont {TABLE I.} List of Selected Radio Stations\par}
    \vspace{4pt} % optional spacing between caption and image
    \includegraphics[width=0.8\linewidth]{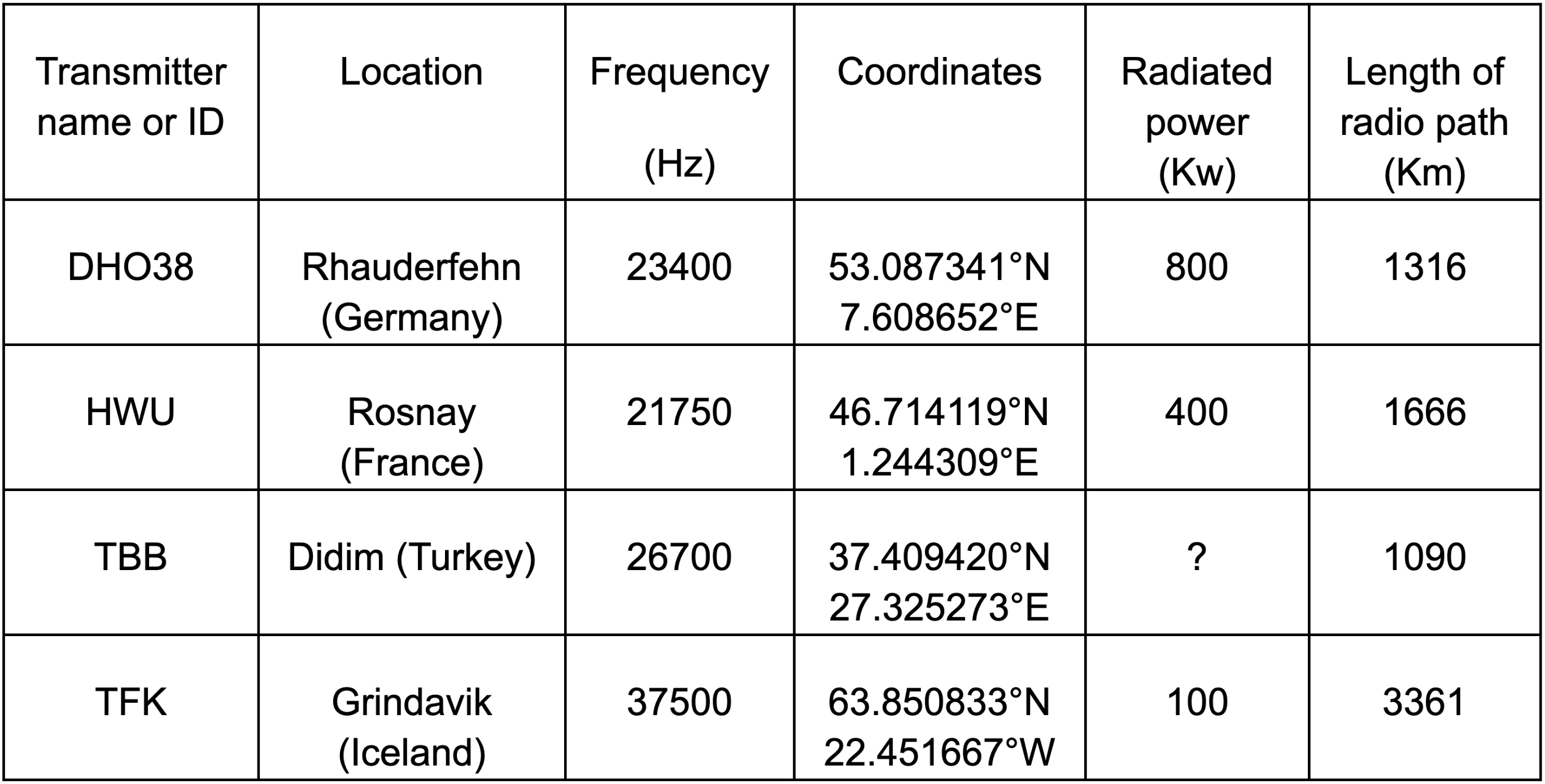}
    \label{fig:selected-stations}
\end{figure}

\begin{figure}[H]
    \centering
    \includegraphics[width=0.75\linewidth]{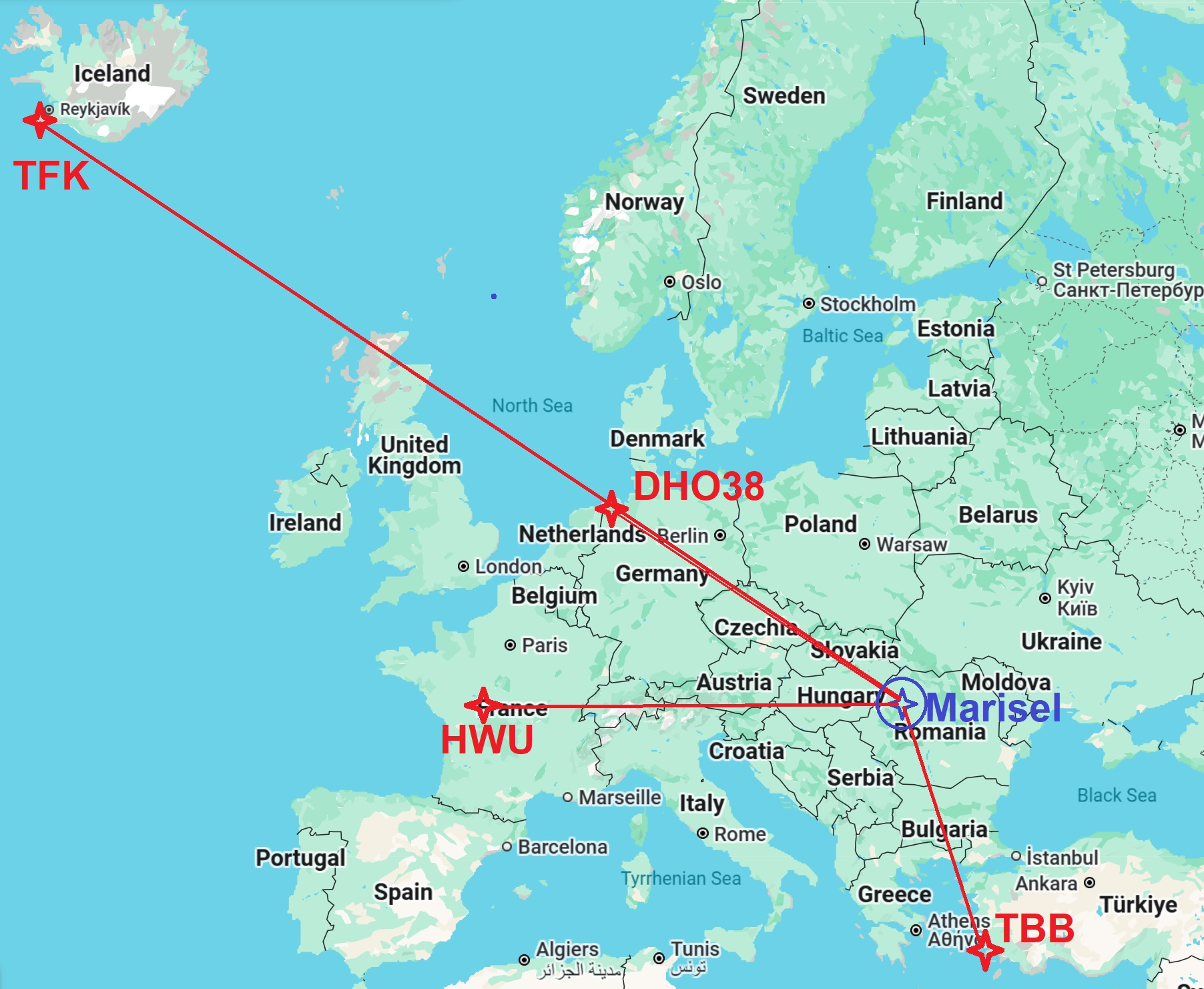}
    \caption{\fontsize{8}{10}\selectfont Map of Selected VLF Transmitters}
    \label{fig:enter-label}
\end{figure}

\section{Experimental Results}

\subsection{Signal Reception}

In Spectrum Lab, several peaks are detected between 19 kHz and 25 kHz, corresponding to signals from nearby VLF transmitters in Europe. These signals are receivable at our location, with the strongest ones shown in Fig. 14. Additional details on radio stations can be found in TABLE II.

\begin{figure}[H]
    \centering
    \includegraphics[width=0.85\linewidth]{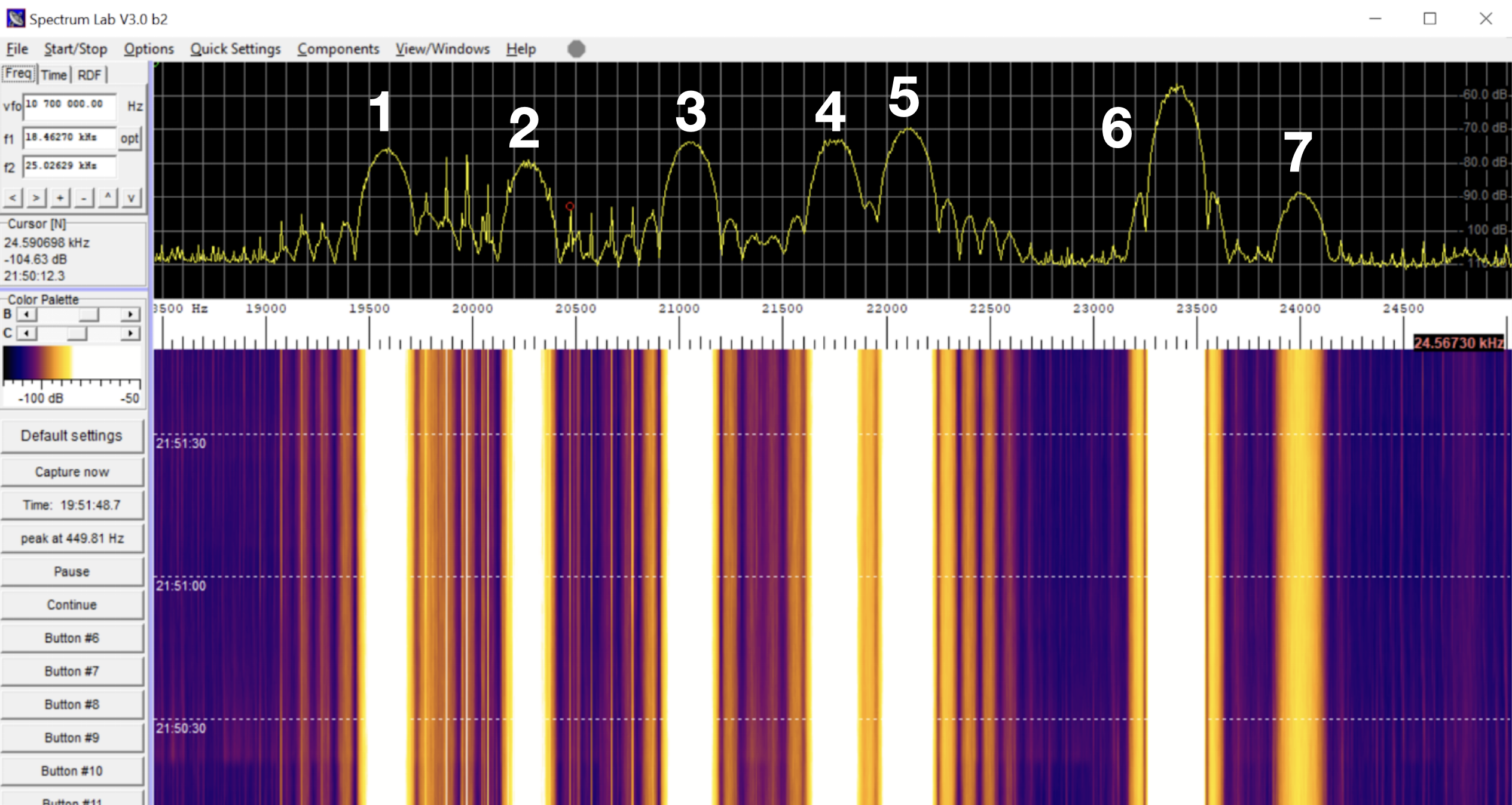}
    \caption{\fontsize{8}{10}\selectfont VLF Radio peaks in Spectrogram}
    \label{fig:enter-label}
\end{figure}

\begin{figure}[H]
    \centering
    {\fontsize{8}{10}\selectfont {TABLE II.} Information about VLF Stations in Fig. 14\par}
    \vspace{4pt}
    \includegraphics[width=1\linewidth]{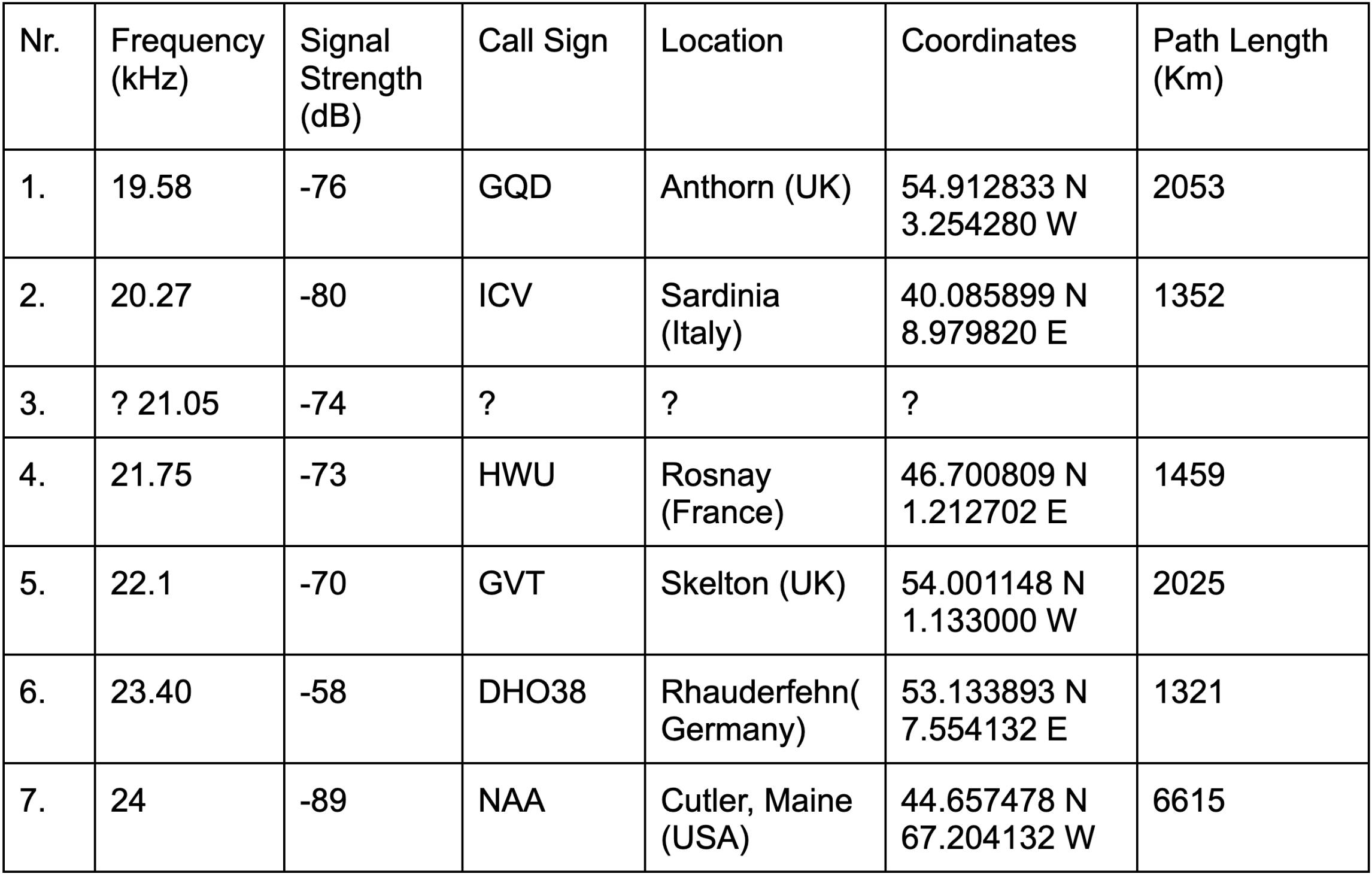}
    \label{fig:vlf-info}
\end{figure}

\subsection{Amplitude vs Time Measurements}

Data was gathered from Marisel, Romania ( 46.659 N, 23.109 E), processed and plotted over the course of several weeks. This location was chosen as it is located in a remote, electromagnetically quiet zone. This is required for the system for it cannot operate properly near urban settlements. Results are presented in Figs. 15 through 17 :

\begin{figure}[H]
    \centering
    \includegraphics[width=1\linewidth]{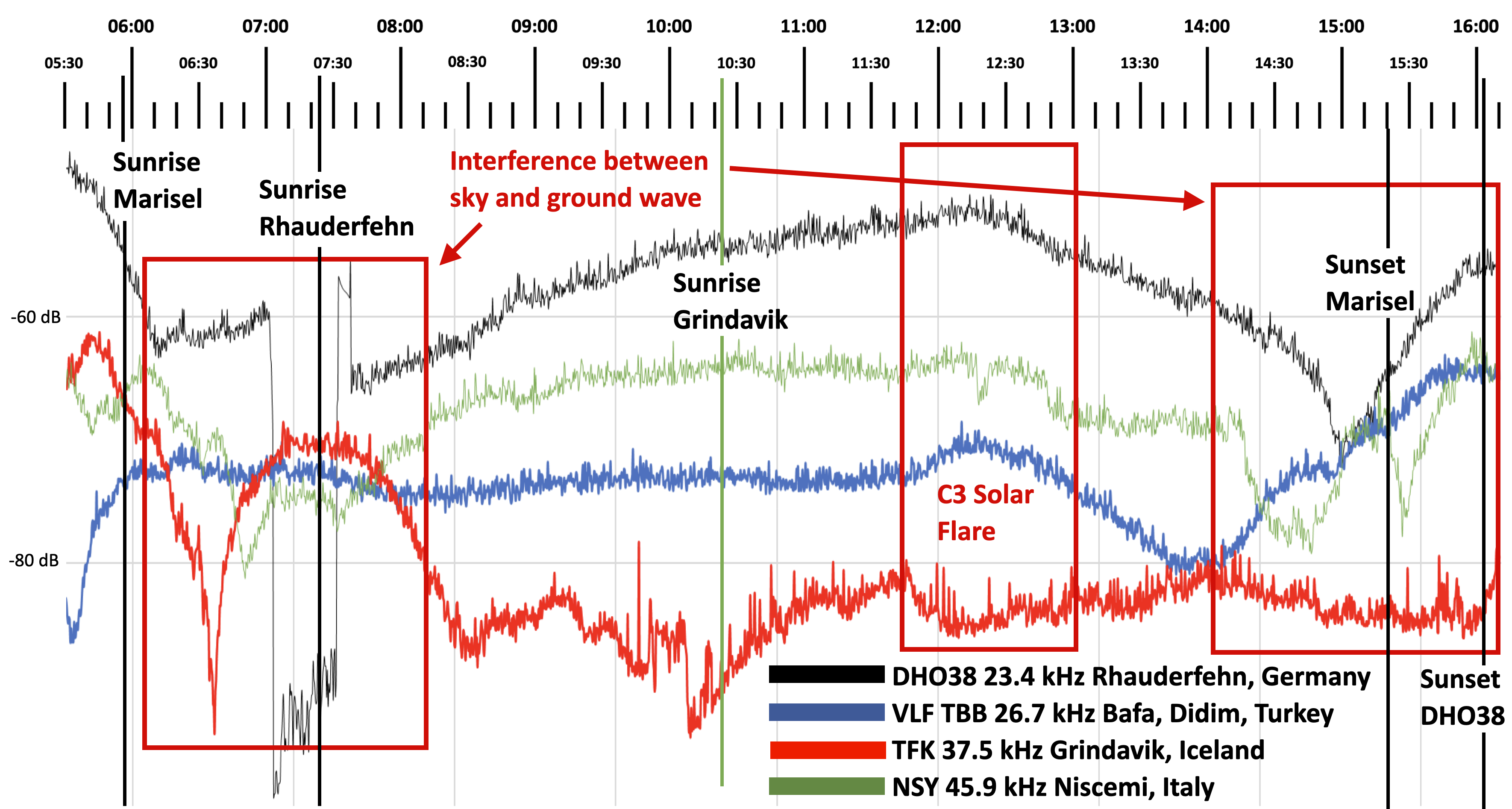}
    \caption{\fontsize{8}{10}\selectfont VLF Data on 24.01.2025}
    \label{fig:enter-label}
\end{figure}

\begin{figure}[H]
    \centering
    \includegraphics[width=1\linewidth]{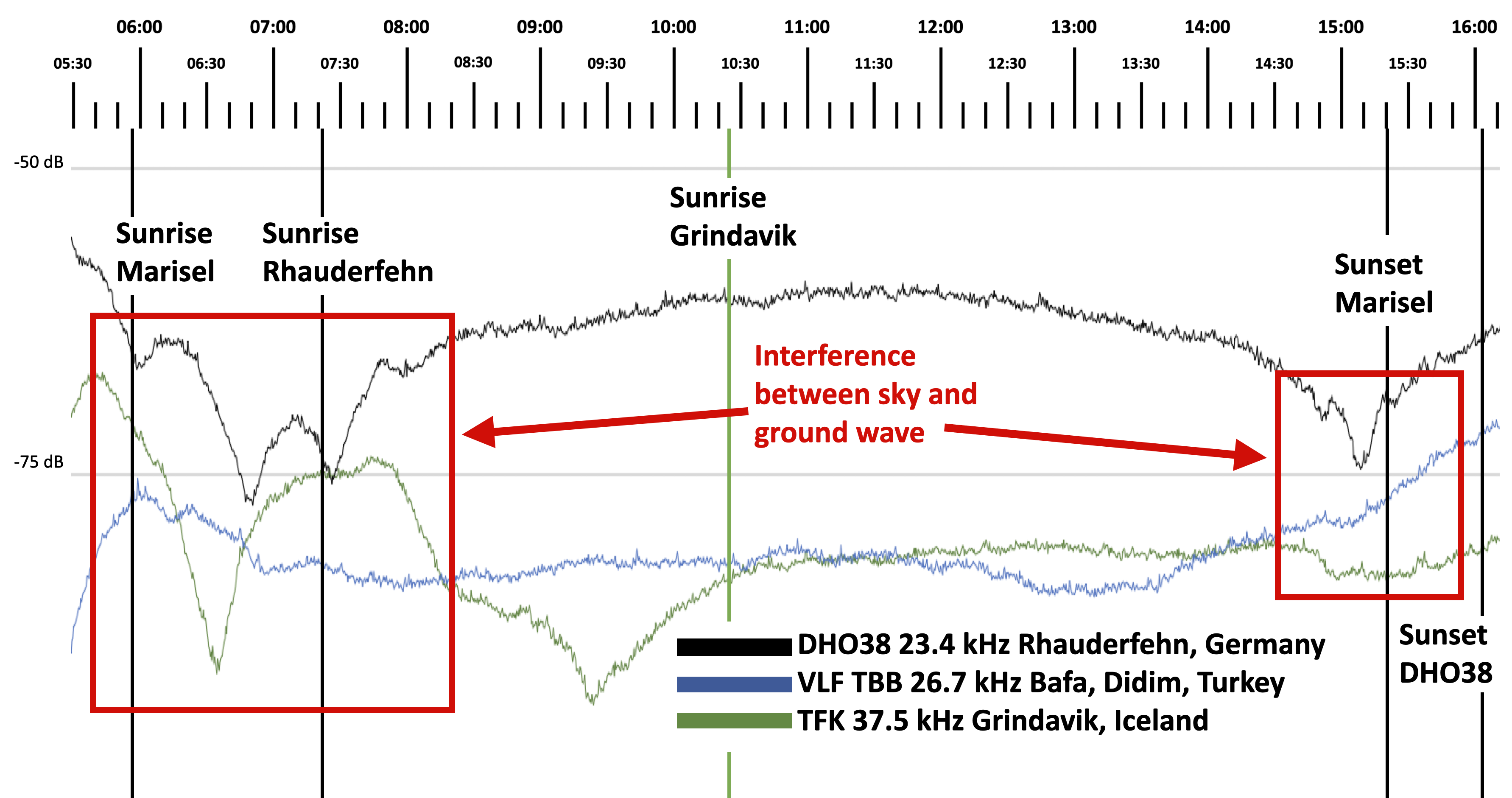}
    \caption{\fontsize{8}{10}\selectfont VLF Data 26.01.2025}
    \label{fig:enter-label}
\end{figure}

\begin{figure}[H]
    \centering
    \includegraphics[width=1\linewidth]{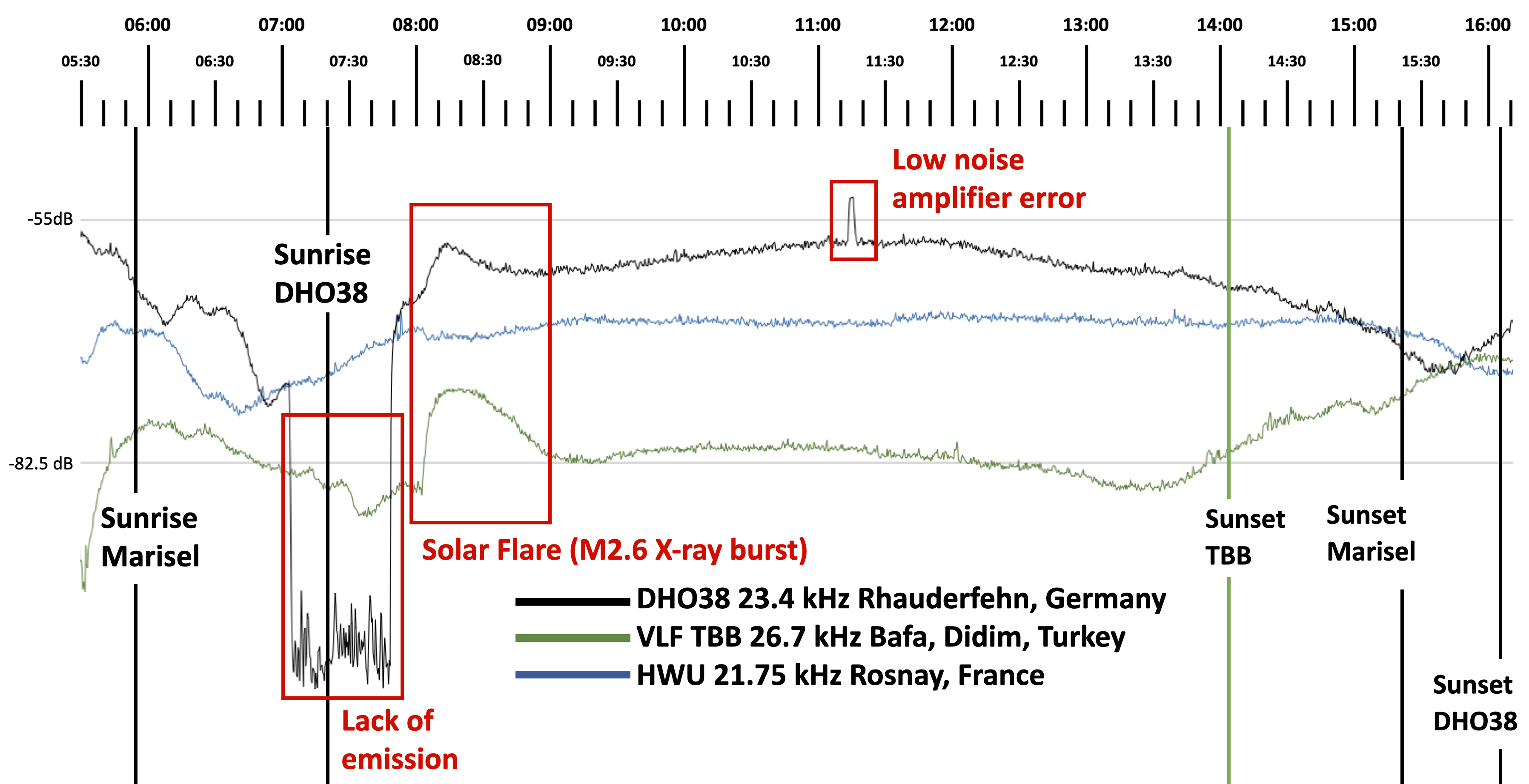}
    \caption{\fontsize{8}{10}\selectfont VLF Data 27.01.2025}
    \label{fig:enter-label}
\end{figure}

See sunrise/sunset times around the world here : \cite{DateandTime}

In Fig 16, in the early hours of the morning we can see multiple peaks and dips in signal intensity, corresponding to constructive/destructive interference between the sky wave and ground wave as they shift past each other, a direct result of the ionosphere changing in thickness. In Fig 17, roughly between 07:00 UTC and 08:00 UTC the signal from DHO38 drops to background noise levels. This is the transmitter turning off, presumably due to maintenance. At 08:00-08:05 UTC we see a sharp rise in signal on all three radio stations, followed by a gradual decrease, as charged particles in the ionosphere recombine. This perfectly coincides with data measured by the GOES satellite \cite{laspGoesXrayFlux} and the sharp rise can be certainly attributed to a class M2.6 solar flare (See Figure 19 ). We can see similar results in Fig 15 : At roughly 12:30 PM there is a slight change in signal strength over all 4 stations that coincides with a class C3 solar flare as measured by the GOES X-ray satellite (see Fig 18). This is the weakest solar event that we have managed to capture with our setup.

\begin{figure}
    \centering
    \includegraphics[width=1\linewidth]{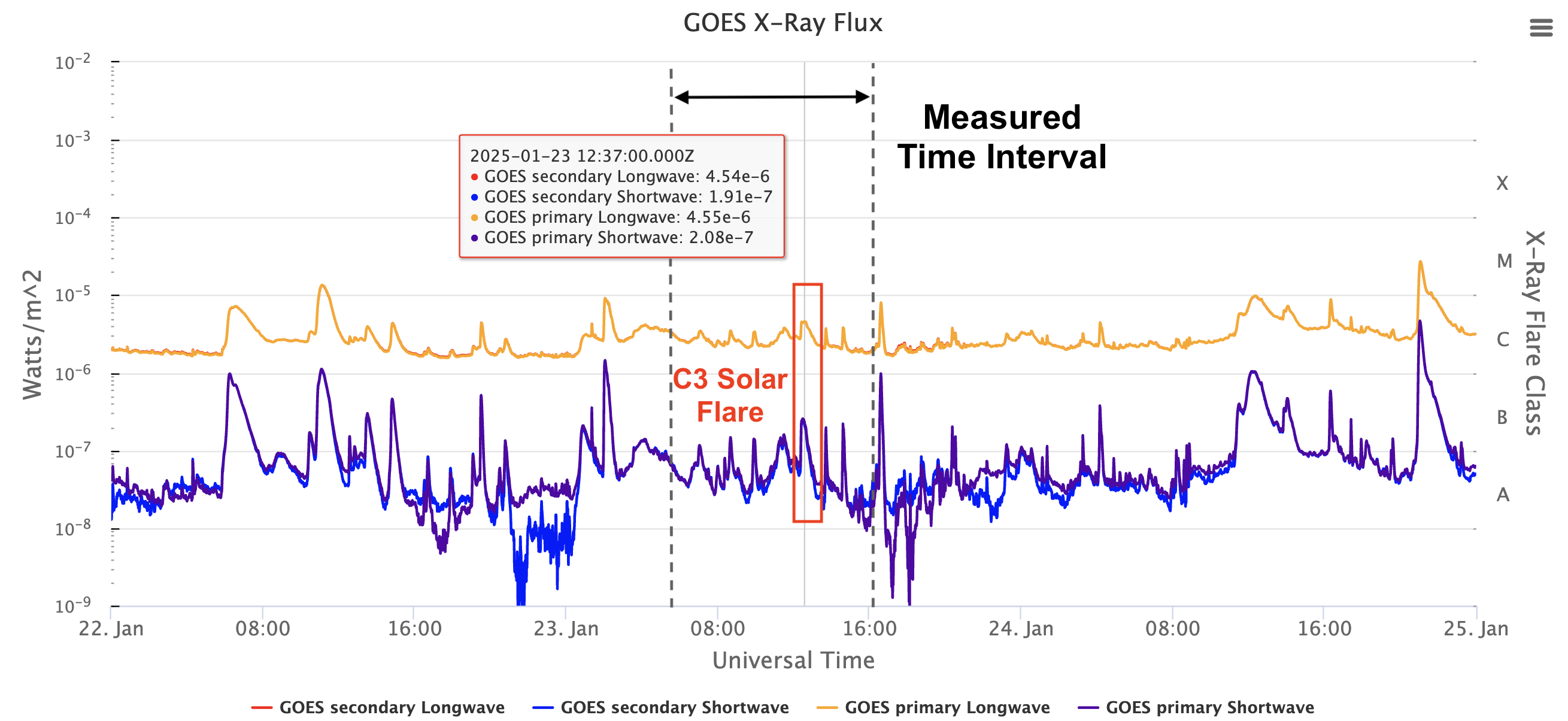}
    \caption{GOES X-ray Flux Data on 24.10.2025}
    \label{fig:enter-label}
\end{figure}

\begin{figure}[H]
    \centering
    \includegraphics[width=1\linewidth]{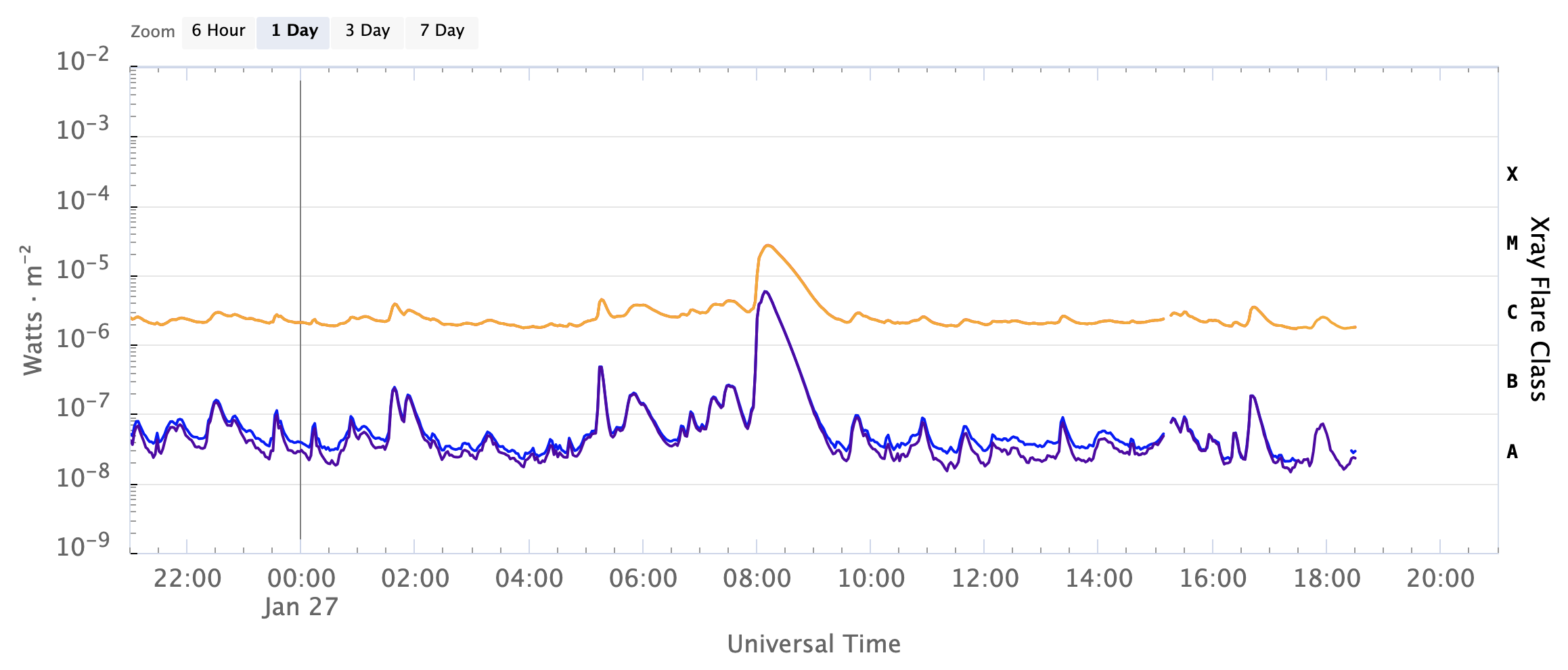}
    \caption{\fontsize{8}{10}\selectfont GOES X-ray Flux Data on 27.01.2025 \cite{laspGoesXrayFlux}}
    \label{fig:enter-label}
\end{figure}

\section{Discussion}

Our findings confirm that the proposed method effectively detects solar flares and signal fluctuations occurring during sunrise and sunset. The observed solar flare profiles, distinctly illustrate the sudden ionization and subsequent recombination of charged particles within the ionosphere. \cite{dolea2013situ} 

Furthermore, due to the longitudinal positioning of the measurement sites — Bafa being the easternmost, followed by Rhauderfehn, and Rosnay the westernmost — the angle between the solar radiation and radial direction from the transmitter site to the earth's center, in the early morning hours, increases respectively across these locations. This results in greater flux variations at more eastern sites, as the X-ray flux is strongest where the solar incidence angle is smallest, a relationship directly explained by the cosine dependence of radiation flux. This phenomenon can be perfectly observed in data captured on 27.01.2025. Signal variation due to a class M2.6 solar flare increases from Rosnay to Bafa.

The system has demonstrated its ability to provide early warnings of geomagnetic storms dozens of hours before they impact the Earth's magnetosphere.

Despite its effectiveness, the method has certain limitations. It is inherently directional, meaning the antenna must be aligned with the transmitting VLF station to achieve maximum signal reception, and external radio interference can degrade performance. Additionally, solar flare detection is restricted to daylight hours, as the system cannot observe flares when the receiving station is on the dark side of Earth. Lastly, the method depends on man-made radio transmissions, making it reliant on the continued operation of such stations.

While the last two constraints are intrinsic to the approach, the issue of directionality could be mitigated by employing a dual-antenna setup with elements positioned perpendicularly. This configuration would enable the system to not only measure signal strength but also determine the direction of incoming signals. Future research is necessary to validate the effectiveness of this proposed enhancement.\\

\section{Conclusion}

VLF radio signals are a reliable, cheaper and convenient Earth-based way of detecting solar flares, which can be useful to predict geomagnetic storms (created by strong solar flares). The system is more accessible to independent researchers ad can be built by someone with knowledge in the field.\\

\section*{Acknowledgment}

I would like to thank Dr. Paul Dolea for his guidance and inspiration in writing this paper.\\

     % basic style, author-year citations
  % name your BibTeX data base
\bibliographystyle{IEEEtran}
\bibliography{references}
\end{document}